\documentclass[
	twocolumn,
	reprint,
	amsmath,
	amssymb,
	aps,
]{revtex4-2}

\usepackage{graphicx} 
\usepackage{dcolumn} 
\usepackage[T1]{fontenc}
\usepackage{mathptmx}
\usepackage{xcolor}

\begin{document}

\title{Spatial locking of chimera states to frequency heterogeneity in nonlocally coupled oscillators}

\author{Petar Mircheski\footnote{Corresponding author: mircheski.p.aa@k.sc.e.titech.ac.jp}}
\affiliation{Department of Systems and Control Engineering, Institute of Science Tokyo, Tokyo 152-8552, Japan}

\author{Hiroya Nakao}
\affiliation{Department of Systems and Control Engineering, Institute of Science Tokyo, Tokyo 152-8552, Japan}

\date{\today}

\begin{abstract}
	Chimera states in systems of nonlocally coupled oscillators, i.e., self-organized coexistence of coherent and incoherent oscillator populations, have attracted much attention.
	In this study, we consider the effect of frequency heterogeneities on the chimera state and reveal that it induces spatial locking of the chimera state, i.e., the coherent and incoherent domains align with
	lower and higher frequency regions, respectively, in a self-adaptive manner.
	Using an extended self-consistency approach, we show that such spatially locked chimera states can be reproduced as steady solutions of the system in the continuum limit.
	Furthermore, we develop a variational argument to explain the mechanism leading to spatial locking.
	Our analysis reveals how heterogeneity can affect the collective dynamics of the chimera states and offers insights into their control and applications.
\end{abstract}

\maketitle

Chimera states observed in networks of coupled oscillators are intriguing spatiotemporal phenomena in which synchronized (coherent) and unsynchronized (incoherent) oscillators coexist even though the properties of the oscillators are identical.
They were first reported by Kuramoto~and~Battoghtogh~\cite{kuramoto2002coexistence} on a ring of nonlocally coupled phase oscillators and named chimera states by Abrams~and~Strogatz~\cite{abrams2004chimera}.
It is also worth noting that the coexistence of coherent and incoherent populations had been studied even earlier in nonlocally or globally coupled oscillators by Kuramoto~\cite{kuramoto1995scaling, kuramoto1997power} and others~\cite{nakao1999anomalous,nakagawa1993collective, chabanol1997collective} in different setups.

Over the past two decades, chimera states have attracted much interest of research.
They have been studied under different types of coupling schemes~\cite{kuramoto2002coexistence, omel2010chimera, abrams2004chimera}, oscillator dynamics~\cite{kuramoto2002coexistence, hizanidis2014chimera, zur2018chimera}, both in one-dimensional and multidimensional systems~\cite{maistrenko2015chimera, omel2012stationary}.
In experiments, chimera states have been observed in various types of systems such as chemical oscillators~\cite{tinsley2012chimera, wickramasinghe2013spatially}, mechanical oscillators~\cite{martens2013chimera, kapitaniak2014imperfect}, and optical oscillators~\cite{hagerstrom2012experimental}.
Effects of external perturbations on the chimera states have also been studied, such as adjusting the spatial positions of the chimera states, controlling the lifetime of the chimera states, and stabilizing the chimera states~\cite{bick2015controlling, isele2016controlling, gambuzza2016pinning, omelchenko2016tweezers, ruzzene2019controlling, yao2019self}.

In this study, we consider chimera states on a ring array of phase oscillators in the continuum limit, which is the standard setup in the theoretical analysis~\cite{kuramoto2002coexistence,abrams2004chimera}.
If the oscillators are homogeneous, the positions of the steady coherent and incoherent domains are not fixed and are determined by the initial conditions due to the continuous translational symmetry.
We show that, when we introduce small frequency heterogeneity to the oscillators, the coherent and incoherent domains are attracted to the lower-frequency and higher-frequency regions, respectively.
That is, {\it the chimera state is spatially locked to the applied frequency heterogeneity in a self-adaptive manner}.
Through the use of the self-consistency approach, we theoretically explain the general mechanism leading to the self-adaptation of the chimera states to frequency heterogeneities.

We consider a spatially extended system of continuously distributed phase oscillators with nonlocal coupling on a ring~\cite{kuramoto2002coexistence,abrams2004chimera}, described by
\begin{equation}
	\dot{\theta}(x, t) = \omega(x) + \int_{-\pi}^{\pi} G(x-x') \sin(\theta(x', t) - \theta(x, t) - \alpha) d x'.
	\label{eq:hetero-model}
\end{equation}
Here, $\theta(x,t)$ is the phase of the oscillator at time $t$ and position $x$, $\omega(x)$ is the natural frequency of the oscillator at position $x$, and the Kuramoto-Sakaguchi sinusoidal coupling with phase lag $\alpha$ is assumed.
We define the spatial domain as $-\pi \leq x \leq \pi$ and assume periodic boundary conditions.
The coupling kernel $G(x-x')$ is a non-negative even function that defines the nonlocal coupling between the oscillators.
It is generally assumed that all oscillators are identical, i.e., $\omega(x) \equiv \omega_0$ with a constant frequency $\omega_0$, where we can set $\omega_0=0$ without loss of generality.
\begin{figure}[htbp]
	\centering
	\includegraphics[width=\hsize]{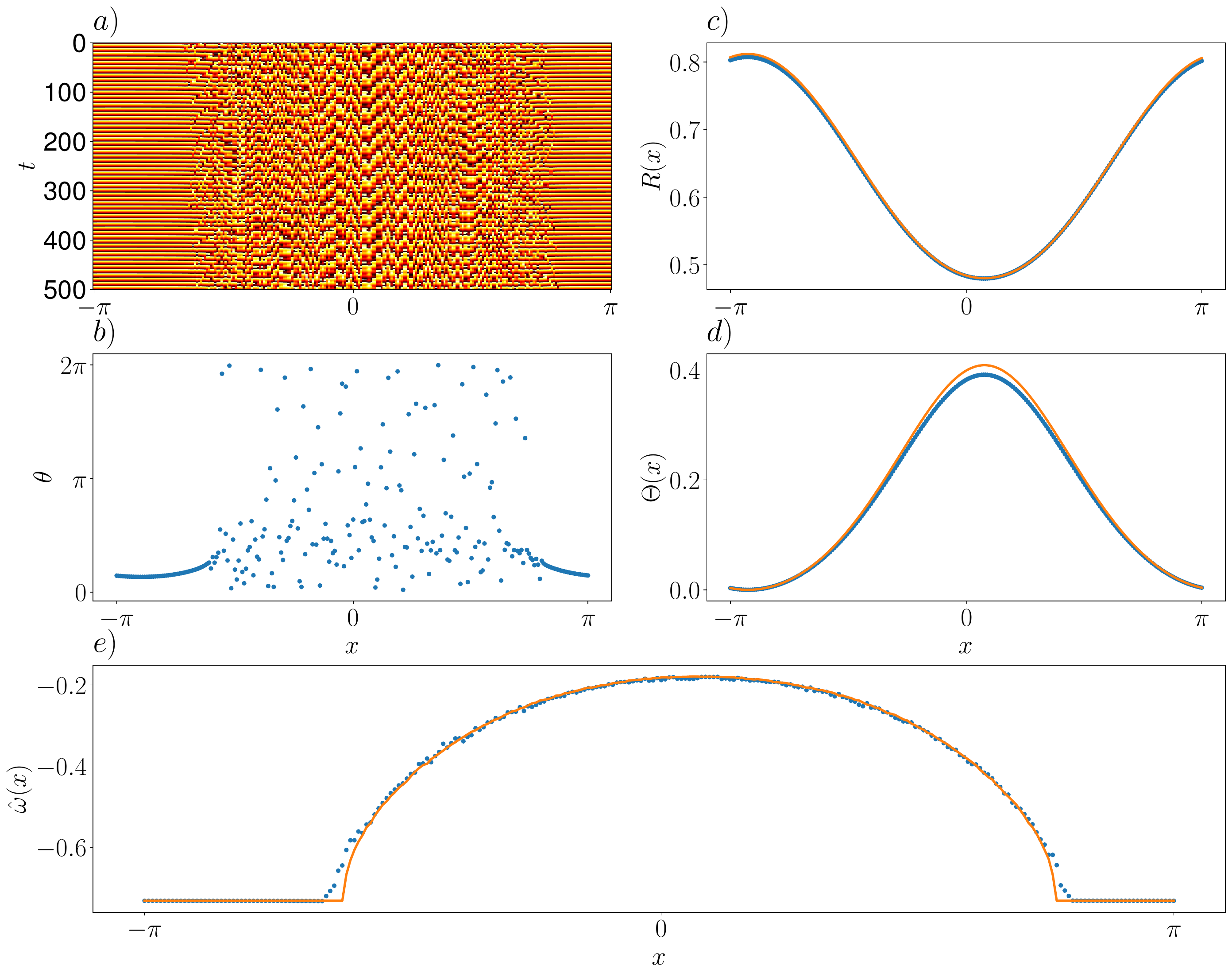}
	\caption{(a) Time evolution of typical phase patterns in the chimera state obtained by simulating Eq.~\eqref{eq:hetero-model} coupled through the cosine kernel, in the absence of heterogeneity ($\epsilon=0$). (b) Snapshot of the phase pattern at $t=500$. (c) Local phase coherence $R(x)$. (d) Local average phase $\Phi(x)$. e) Instantaneous frequency $\hat{\omega}(x)$. Blue points show the results of numerical simulations, and the orange curves show the numerical solutions of the self-consistency equation~\eqref{eq:condition}. The initial conditions are sampled from $\theta(x, 0) = 6 e^{-0.76x^2} r(x)$ where $r(x)$ is a uniform random variable in the interval $-1/2 < r(x) < 1/2$.}
	\label{fig:typical-pattern}
\end{figure}

From now on, we assume that the kernel has a cosine shape as in~\cite{abrams2004chimera} rather than the exponential shape as in~\cite{kuramoto2002coexistence},
\begin{equation}
	G(x-x') = (1 + A \cos(x - x')) \frac{1}{2\pi},
	\label{eq:cosine-kernel}
\end{equation}
and fix the system parameters to $\alpha = \frac{\pi}{2} -0.18$ and $A = 0.995$.
This is the simplest setting in which the model displays the well-researched chimera states.
Even if the system is homogeneous, $\omega(x) \equiv \omega_0$, it can self-organize into coexisting coherent and incoherent domains, i.e., chimera states, as shown in Figs.~\ref{fig:typical-pattern}(a) and ~\ref{fig:typical-pattern}(b).
In numerical simulations, the system is discretized using $N=256$ grid points (i.e., oscillators) and the integration is approximated by a finite sum.
The resulting phase patterns of the chimera states have been analyzed by using the self-consistency analysis~\cite{kuramoto2002coexistence,abrams2004chimera} as explained below.
As the system is spatially homogeneous, the spatial position of the resulting chimera state is arbitrary and determined by the initial condition.
We take the center of the self-organized incoherent domain at $x=0$ without loss of generality.

\begin{figure}[htbp]
	\centering
	\includegraphics[width=\hsize]{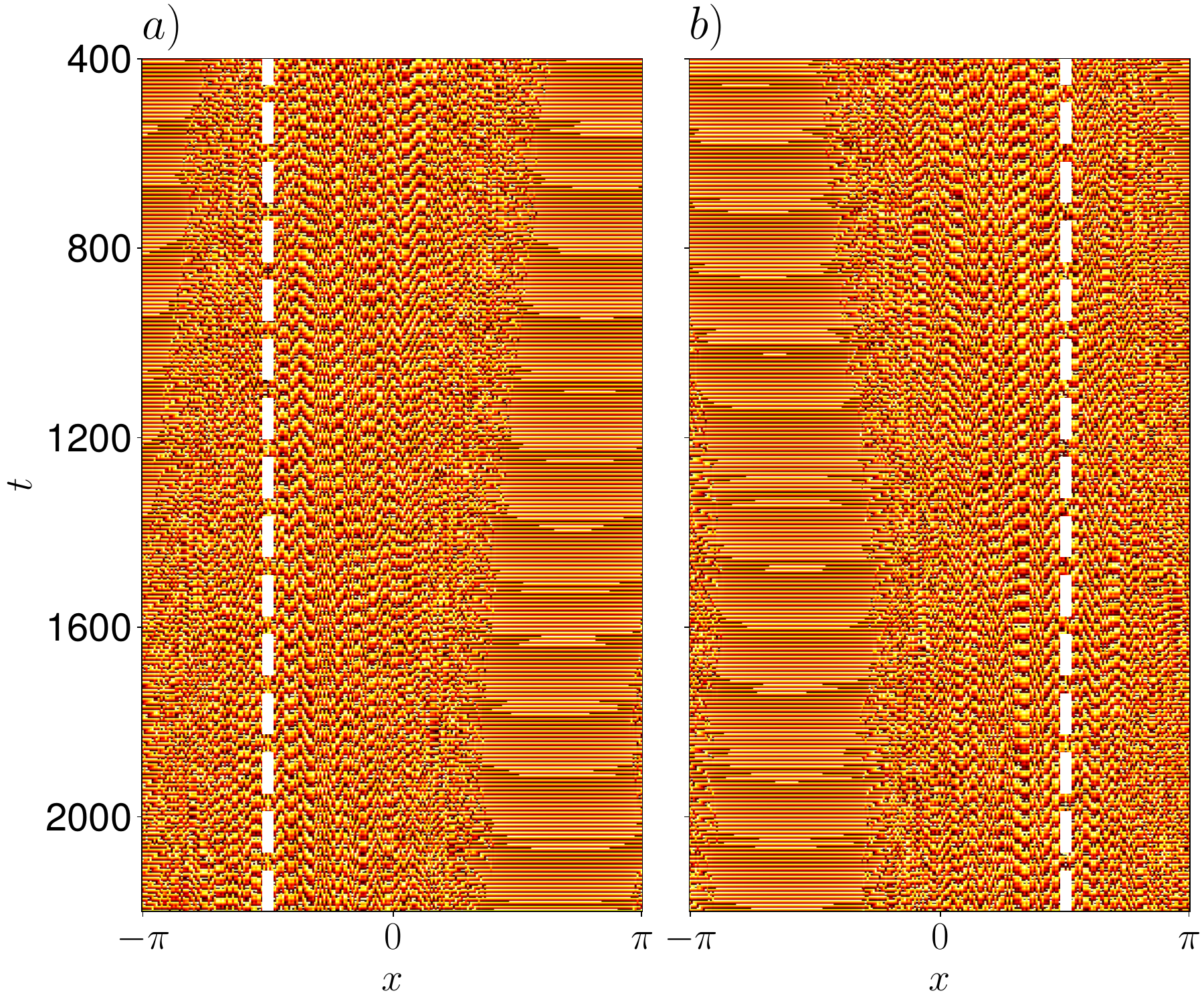}
	\caption{Evolution of phase patterns obtained by simulating Eq.~\eqref{eq:hetero-model} with the point heterogeneity. As the initial condition, we take the steady chimera state in Fig~\ref{fig:typical-pattern}. The white dashed line gives the location $\gamma$ of the point heterogeneity. (a) $\gamma=-\pi/2$ and (b) $\gamma=\pi/2$. }
	\label{fig:dirac-movement}
\end{figure}

\begin{figure}[htbp]
	\centering
	\includegraphics[width=\hsize]{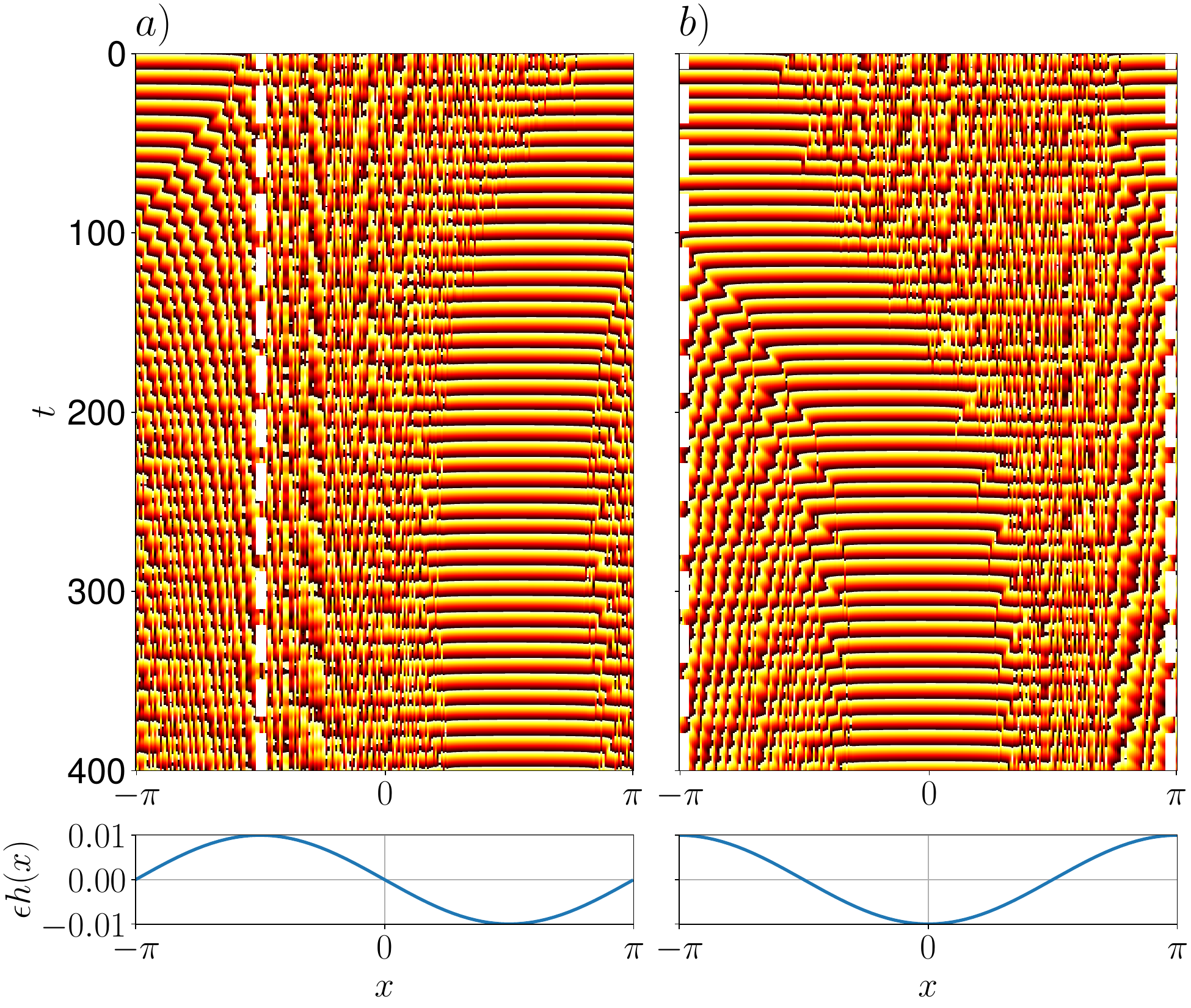}
	\caption{Evolution of phase patterns under the cosine heterogeneity obtained by simulating Eq.~\eqref{eq:hetero-model}. The steady chimera state in Fig~\ref{fig:typical-pattern} is taken as the initial condition. The white dashed line indicates the location $\gamma$ of the cosine heterogeneity. (a) $\gamma=-\pi/2$ and (b) $\gamma = \pi$.}
	\label{fig:cosine-movement}
\end{figure}

In this study, we consider the effect of weak spatial frequency heterogeneity on the chimera state.
Thus, we assume
\begin{equation}
	\omega(x) = \omega_0  + \epsilon h(x-\gamma)
\end{equation}
in Eq.~\eqref{eq:hetero-model}, where $\epsilon$ ($0 < \epsilon \ll 1$) is a small parameter characterizing the magnitude of the frequency heterogeneity, $h(x)$ is a periodic even function defined on $-\pi \leq x \leq \pi$ characterizing the spatial shape of the frequency heterogeneity, and
$-\pi \leq \gamma \leq \pi$ is a free parameter that defines the location of the heterogeneity on the ring.
We assume that $\epsilon$ is sufficiently small so that the chimera states are only slightly perturbed.
We consider two types of frequency heterogeneities, i.e., a Dirac-like point heterogeneity and a smooth cosine heterogeneity, as two typical examples.
We stress that similar results are observed generally for other localized frequency heterogeneities.

\begin{figure}[htbp]
	\centering
	\includegraphics[width=\hsize]{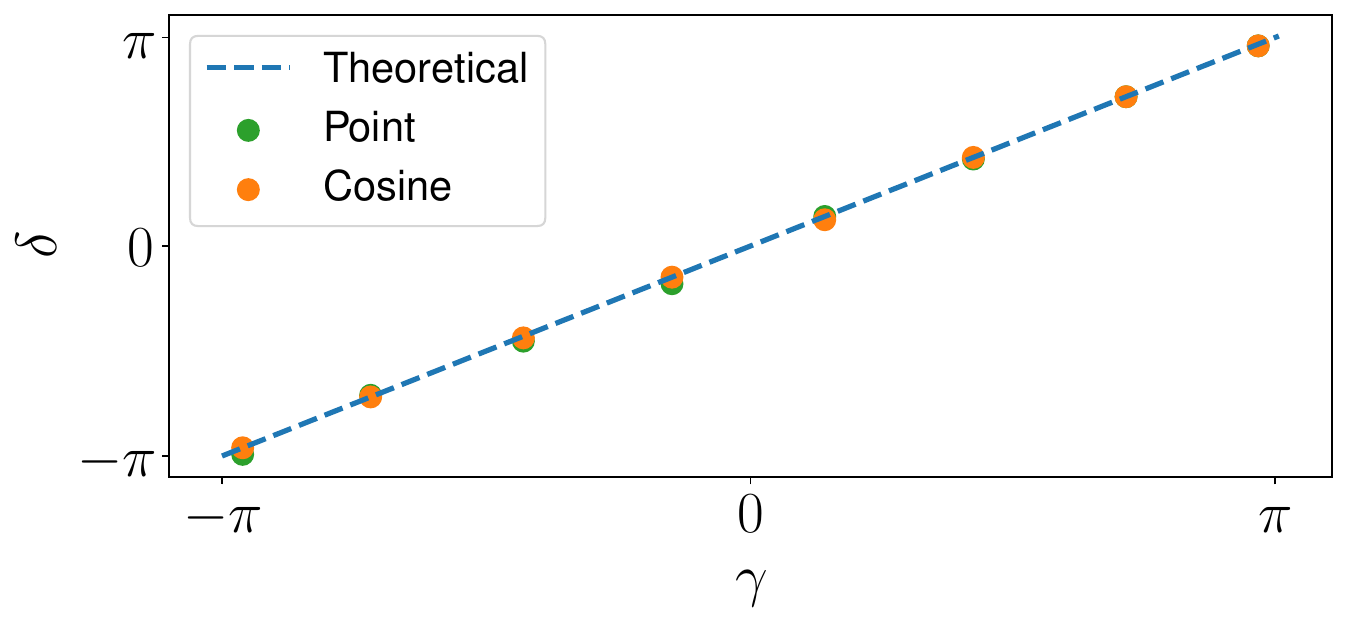}
	\caption{Position $\delta$ of the incoherent domain of the chimera state, with respect to the offset of the heterogeneity $\gamma$. The dashed line shows the theoretical position of the center of the incoherent domain, while orange and green markers display the position of the center under point heterogeneity and cosine heterogeneity, respectively.}
	\label{fig:prediction-dots}
\end{figure}

For the point heterogeneity, we assume
\begin{equation}
	h(x) = \tilde\delta(x),
	\label{eq:dirac-hetero}
\end{equation}
where $\tilde\delta(x)$ is Dirac's delta function~\footnote{In the numerical simulation, this is replaced by a frequency increase at a single grid point in the system, i.e., $\omega_0 \to \omega_0 + \epsilon / \Delta x$ at the grid corresponding to $x = \gamma$, where $\Delta x = 2\pi / N$ is the grid width.}.
We set the magnitude of heterogeneity as $\epsilon=0.2$.
We numerically simulate Eq.~\eqref{eq:hetero-model} from the initial chimera state of the homogeneous system shown in Fig.~\ref{fig:typical-pattern} and apply the heterogeneity at two different locations ($\gamma = \pm \pi/2$) from $t=0$.
Figure~\ref{fig:dirac-movement} displays the chimera states under the effect of this point frequency heterogeneity obtained by numerical simulations.
We can observe that the chimera states are maintained under the heterogeneity.
Moreover, the coherent domain drifts and settles in a region far away from the point heterogeneity, while the incoherent domain is attracted to the location of the point heterogeneity, as shown in  Fig.~\ref{fig:prediction-dots}, where the position $\delta$ of the center of the incoherent domain in the steady state is plotted with respect to $\gamma$. Thus, the higher-frequency oscillator attracts the incoherent domain and repels the coherent domain.

For the cosine heterogeneity, we assume
\begin{equation}
	h(x) = \cos(x)
\end{equation}
and set $\epsilon = 0.01$.
Figure~\ref{fig:cosine-movement} shows the results of numerical simulations for $\gamma=-\pi/2$ and $\gamma=\pi$.
Similar to the case with the point heterogeneity, the chimera state is maintained, and the region of higher-frequency oscillators repels the coherent domain while attracting the incoherent domain.
Eventually, the coherent and incoherent domains settle perfectly at the positions of the lowest and highest natural frequencies, respectively.
In Fig.~\ref{fig:prediction-dots}, the center $\delta$ of the coherent domain in the steady state is plotted versus $\gamma$ also for the case with the cosine heterogeneity, clearly confirming this phenomenon.

The above results suggest that, despite the different forms of heterogeneity, the chimera states exhibit the same behavior:
{the coherent domain drifts away from the region where the oscillators have higher natural frequencies and eventually aligns with regions of lower natural frequencies (and vice versa for the incoherent domain).}
This consistent behavior highlights the system's simple and fundamental dynamics of self-adaptation with respect to the
weak frequency heterogeneity, i.e., {\it spatial phase locking to frequency heterogeneity}.

\begin{figure}[htbp]
	\centering
	\includegraphics[width=\hsize]{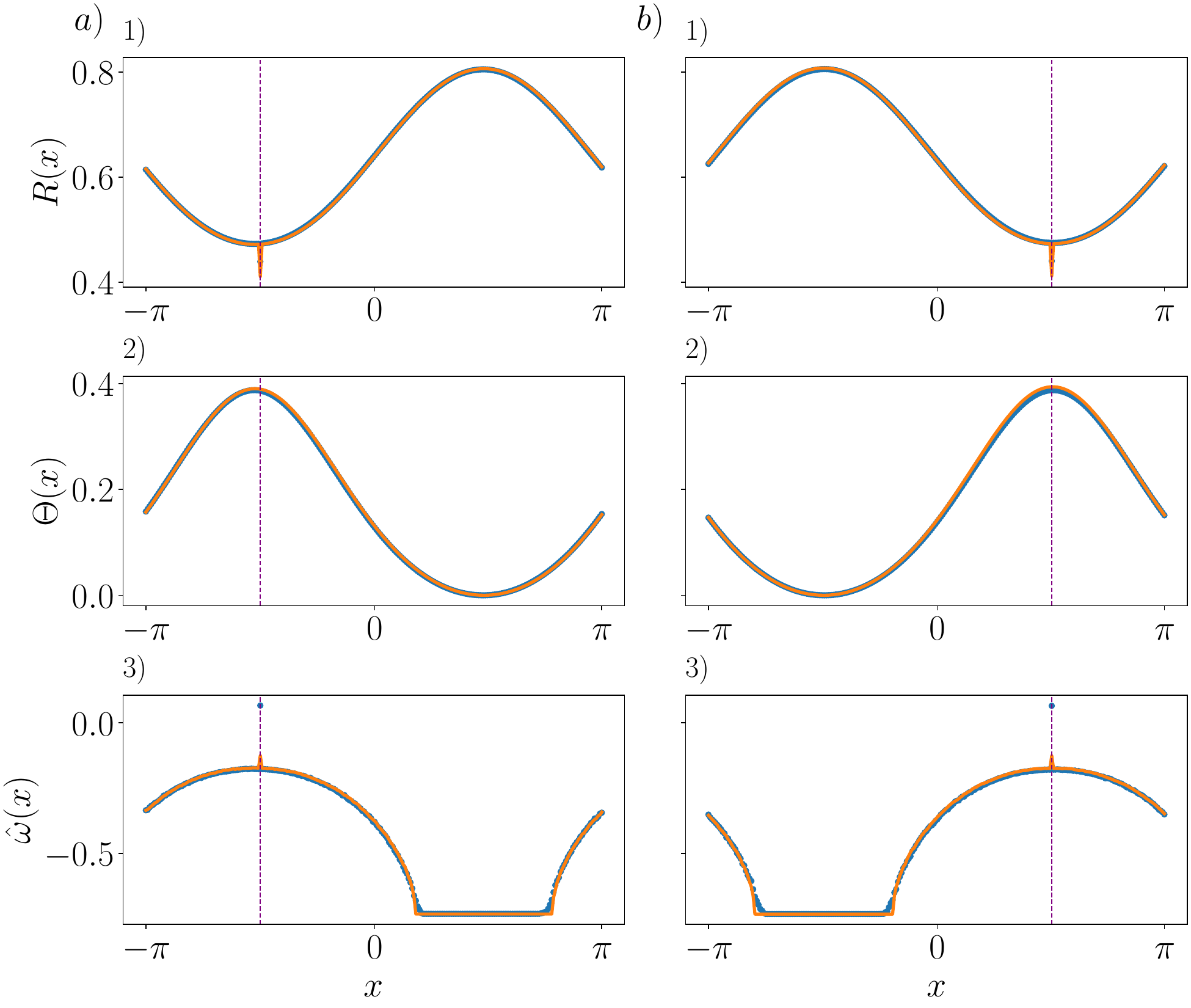}
	\caption{Local order parameters in the steady chimera state under the point heterogeneity with parameters as in Fig~\ref{fig:dirac-movement}. Panels (a) for $\gamma=-\pi/2$ and (b) for $\gamma=\pi/2$ include (1)~Local phase coherence $R(x)$, (2)~Local collective phase $\Phi(x)$, and (3)~Instantaneous frequency $\hat{\omega}(x)$. Blue points show the results of numerical simulations of Eq.~\eqref{eq:hetero-model}, and the orange curves show the solutions of the self-consistency equation~\eqref{eq:condition}. The purple dashed line displays the location of the point heterogeneity.}
	\label{fig:dirac-movement-self-consistency}
\end{figure}

We now try to explain this characteristic spatial locking behavior theoretically by using self-consistency analysis.
To this end, we generalize the previous self-consistency theories for chimera states in~\cite{kuramoto2002coexistence,abrams2004chimera},
originally for homogeneous oscillators, to Eq.~\eqref{eq:hetero-model} with the frequency heterogeneity.
First, we rewrite Eq.~\eqref{eq:hetero-model} with respect to a reference frame with a collective frequency $\Omega$ by introducing $\phi(x,t) = \theta(x,t) - \Omega t$.
Next, by incorporating the frequency heterogeneity $\epsilon h(x)$, we introduce a modified complex local order parameter $A$ and its modulus $R$ and argument $\Phi$ as
\begin{align}
	A(x,t) = R(x,t) e^{i \Phi(x, t)} = \int_{-\pi}^{\pi} G(x-x') e^{i \phi(x', t)} d x' \nonumber \\
	+ \epsilon h(x-\gamma) e^{i (\phi(x, t) + \alpha + \frac{\pi}{2})}.
	\label{eq:order-param-modified}
\end{align}
When the oscillators are homogeneous ($\epsilon=0$), this goes back to the conventional complex local order parameter in~\cite{kuramoto2002coexistence,abrams2004chimera}.
This definition allows us to rewrite Eq.~\eqref{eq:hetero-model} as
\begin{equation}
	\dot{\phi}(x,t) = \Delta + R(x, t)\sin( \Phi(x, t) - \phi(x, t) - \alpha),
	\label{eq:order-param-system}
\end{equation}
where we introduced the relative frequency $\Delta = \omega_0 - \Omega$.
Thus, we can formally obtain a one-body phase equation with homogeneous $\Delta$ that assumes an identical expression to the previous studies~\cite{kuramoto2002coexistence,abrams2004chimera}.
The oscillator's dynamics at each point $x$ is formally expressed as a one-body problem driven by the nonlocal mean fields $R(x,t)$ and $\Phi(x,t)$.
These $R$ and $\Phi$ are in turn determined by all the oscillators and the frequency heterogeneity in a self-consistent manner.

In the steady chimera state, by choosing an appropriate reference frequency $\Omega$, the local order parameter can be made time-independent.
Thus, assuming that $R$ and $\Phi$ depend only on $x$, we look for the self-consistent stationary solution to Eqs.~\eqref{eq:order-param-modified} and~\eqref{eq:order-param-system} as well as the appropriate value of $\Omega$.
The coherent domain corresponds to the spatial region where $R(x)$ takes larger values and the oscillators are phase-locked to the local collective phase $\Phi(x)$, while the incoherent domain corresponds to the spatial region where $R(x)$ is smaller and the oscillators are drifting from the collective phase.

Let us focus on the coherent domain first, which we denote by $D_c$.
The oscillator in this domain ($x \in D_c$) fulfills $R(x) > |\Delta|$ and is phase-locked to the local collective phase $\Phi(x)$. The phase $\phi(x,t)$ of the oscillator approaches a fixed point $\phi_0(x) = \arcsin \left( |\Delta| / R(x) \right) - \alpha  + \Phi(x)$.
The contribution of these phase-locked oscillators to the order parameter is calculated by substituting the fixed point into Eq.~\eqref{eq:order-param-modified} and restricting the integration range to $D_c$, yielding
\begin{align}
	A_c(x) =
	i e^{-i \alpha} \int_{D_c} G(x-x') e^{i \Phi(x')} \frac{i \sqrt{ R^2(x') - \Delta^2} + \Delta}{R(x')} dx' \nonumber \\
	- \epsilon h(x-\gamma) e^{i \Phi(x)} \frac{i\sqrt{ R^2(x) - \Delta^2} + \Delta}{R(x)}
	\label{eq:coherent-contribution}
\end{align}
In contrast, the oscillator with $R(x) < |\Delta|$ is drifting and belongs to the incoherent domain, denoted by $D_i$.
The probability density of the phase of a drifting oscillator is inversely proportional to its phase velocity $|\dot{\phi}(x, t)|$ and is given by $\rho(\phi, x) = \sqrt{\Delta^2 - R^2(x)} / \big( 2 \pi |\Delta + R(x)\sin(\Phi(x) - \phi(x) - \alpha)| )$, and the statistical average of the contribution from this drifting oscillator to the order parameter is given by $\int_{-\pi}^{\pi} e^{i \phi(x)} \rho(\phi, x) d \phi = (i / R(x)) (\Delta  - \sqrt{\Delta^2 - R^2(x)})$.
Substituting this into Eq.~\eqref{eq:order-param-modified} while restricting the integration range to $D_i$, the average contribution from the oscillators in the incoherent domain $D_i$ is obtained as
\begin{align}
	A_i(x) =
	i e^{-i \alpha} \int_{D_i} G(x-x') e^{i \Phi(x')} \frac{\Delta - \sqrt{\Delta^2 - R^2(x')}}{R(x')} d x' \nonumber \\
	- \epsilon h(x-\gamma) e^{i \Phi(x)} \frac{\Delta - \sqrt{\Delta^2 - R^2(x)}}{R(x)}.
	\label{eq:incoherent-contribution}
\end{align}
Similar to Strogatz~and~Abrams \cite{abrams2004chimera} for the homogeneous case, these contributions of the coherent and incoherent domains are equal; comparing Eq.~\eqref{eq:coherent-contribution} and Eq.~\eqref{eq:incoherent-contribution}, both expressions are identical if we choose the branch of the square root that corresponds to the positive imaginary ($+i$) root when solving for a negative argument inside the square root.
Thus, we can rewrite the complex exponential $e^{i \phi(x)}$ in Eq.~\eqref{eq:order-param-modified} in both domains as
$e^{i {\phi}(x)} = e^{i \beta}  e^{i \Phi(x)} (\Delta - \sqrt{\Delta^2 - R^2(x)})/R(x),$ where we defined $\beta = \frac{\pi}{2} - \alpha$.
Finally, the self-consistency equation for the local order parameter can be written in a concise form,
\begin{align}
	R(x) e^{i \Phi}(x) = e^{i \beta} \int_{-\pi}^{\pi} G(x-x') e^{i \Phi(x')} \frac{\Delta - \sqrt{\Delta^2 - R^2(x')}}{R(x')} d x' \nonumber \\
	- \epsilon h(x-\gamma) e^{i \Phi(x)} \frac{\Delta - \sqrt{\Delta^2 - R^2(x)}}{R(x)}.
	\label{eq:condition}
\end{align}
This equation was solved numerically for the homogeneous case ($\epsilon=0$) in~Fig.~\ref{fig:typical-pattern}, reproducing~\cite{abrams2004chimera}.
Figures~\ref{fig:typical-pattern}c), ~\ref{fig:typical-pattern}d), and ~\ref{fig:typical-pattern}e) compare the results for $R$, $\Phi$, and the instantaneous frequency $\hat{\omega}$ obtained by numerical simulations and by solving the self-consistency equation numerically.

\begin{figure}[htbp]
	\centering
	\includegraphics[width=\hsize]{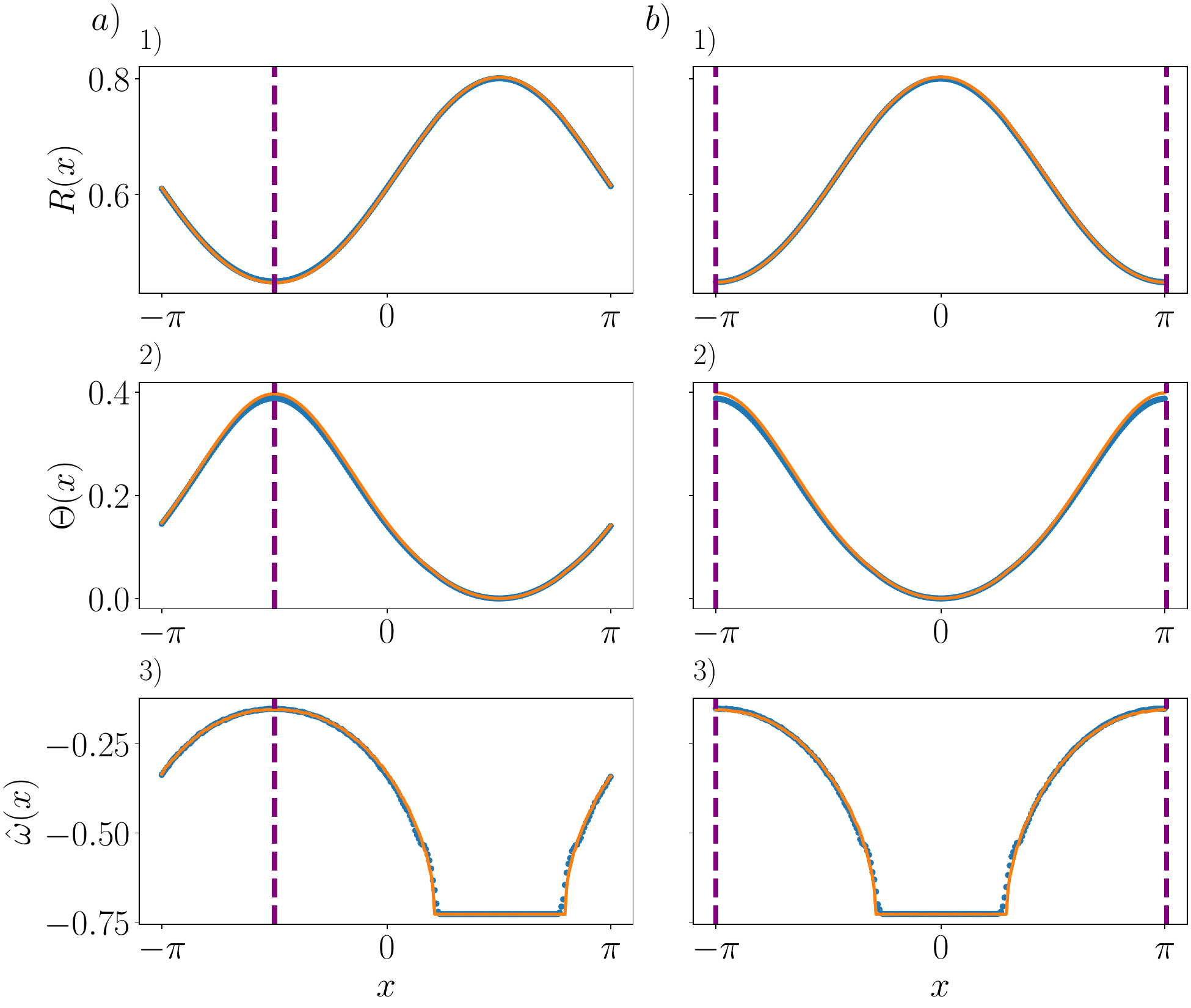}
	\caption{Local order parameters and average instantaneous frequency in the steady chimera state under cosine heterogeneity with parameters as in Fig~\ref{fig:cosine-movement}. The same quantities as in Fig~\ref{fig:dirac-movement-self-consistency} are plotted. (a) $\gamma=-\pi/2$ and (b) $\gamma=\pi$.}
	\label{fig:cosine-movement-self-consistency}
\end{figure}

We now apply the above self-consistency approach to describe the final patterns obtained under spatial heterogeneity.
We focus on the point heterogeneity first.
Figures~\ref{fig:dirac-movement-self-consistency}(a-b) show the self-consistent solution for $R(x)$ and $\Phi(x)$ and the resulting average instantaneous frequency $\hat{\omega}(x)$ obtained numerically for several values of $\gamma$ (location of the heterogeneity).
In the figures, the results obtained by numerical simulations of Eq.~\eqref{eq:hetero-model} are also plotted, which agree well with the self-consistency solutions.
We can clearly confirm that the spatial locking behavior is reproduced in the self-consistency solution.
That is, the coherent domain perfectly positions itself with the opposite side of the ring to avoid the high-frequency oscillator, while the incoherent domain is attracted to the high-frequency oscillator.
Similarly, in the case of cosine heterogeneity,
Figs.~\ref{fig:cosine-movement-self-consistency} (a-b) display the numerical self-consistency solutions for $R(x)$, $\Phi(x)$, and the average instantaneous frequencies $\hat{\omega}(x)$ for several values of $\gamma$. The results of numerical simulations of Eq.~\eqref{eq:hetero-model} are also plotted, showing good agreement.
We can confirm again that the coherent domain prefers the low-frequency region while the incoherent domain is attracted to the high-frequency region.

We have shown that the self-consistency analysis reproduces the general spatial phase-locking behavior of the chimera states under the effect of frequency heterogeneity.
To understand the mechanism determining the final positions of the chimera states, we here develop a variational argument.
First, if no frequency heterogeneity exists ($\epsilon = 0$), we obtain an unperturbed chimera state represented by the unperturbed order parameter given by $R_0(x) e^{i \Phi_0(x)}$, which satisfies the self-consistency equation~\eqref{eq:condition} with $\epsilon = 0$, shown in Fig.~\ref{fig:typical-pattern}. We set the position of the minimum of $R_0(x)$ (center of the incoherent domain) at $x=0$ without loss of generality.

Now, if we introduce weak frequency heterogeneity ($0 < \epsilon \ll 1$), the shapes of $R(x)$ and $\Phi(x)$ remain nearly the same as $R_0(x)$ and $\Phi_0(x)$ in the unperturbed homogeneous case, where the difference is of $O(\epsilon)$ from Eq.~\eqref{eq:order-param-modified}.
At the same time, their spatial positions may shift along the ring.
Such behavior is physically expected as a consequence of the continuous translational symmetry of the homogeneous system on a ring and is indeed confirmed numerically in Figs.~\ref{fig:dirac-movement} and~\ref{fig:cosine-movement}.
Thus, there exists a spatial shift parameter $\delta$ such that the perturbed $R$ and $\Phi$ can approximately be expressed as $R(x) \approx R_0(x - \delta)$, and $ \Phi(x) \approx \Phi_0(x - \delta),$ in the steady chimera state under the frequency heterogeneity.
These are used as trial functions for the variational argument.

The chimera state should satisfy the self-consistency equation~\eqref{eq:condition} also in the heterogeneous case with $\epsilon>0$.
Therefore, we expect that the shift $\delta$ in the above trial functions $R$ and $\Phi$ are chosen so that they satisfy Eq.~\eqref{eq:condition} as much as possible.
Noting that $R_0(x - \delta)$ and $\Phi_0(x - \delta)$ satisfy Eq.~\eqref{eq:condition} with any $\delta$ for $\epsilon=0$, the error is given by the spatial integral of the square of the second term on the right-hand side of Eq.~\eqref{eq:condition},

\begin{align}
	E(\delta)
	 & = \int_{-\pi}^{\pi} \left| \epsilon h(x-\gamma) e^{i \beta} e^{i \Phi_{0}(x - \delta)} \frac{\Delta - \sqrt{\Delta^2 - R^2_0(x-\delta)}}{R_0(x-\delta)}\right|^2 dx \cr
	 & = \epsilon^2 \int_{-\pi}^{\pi} h(x + \delta - \gamma)^2 \left( \frac{\Delta - \sqrt{\Delta^2 - R^2_0(x)}}{R_0(x)}\right)^2 dx,
\end{align}
where the $2\pi$ periodicity of $h$ and $R_0$ are used.
We can assume that $f(x) = \left( (\Delta - \sqrt{\Delta^2 - R^2_0(x)})/R_0(x) \right)^2$ in the integrand has a single minimum at $x=0$, and is an even function $f(x) = f(-x)$, considering the unperturbed chimera in Fig.~\ref{fig:typical-pattern}.
Then, for an even function, $h(x) = h(-x)$, that takes maximum at $x=0$, $E(\delta)$ takes on the minimum at $\delta = \gamma$.
Therefore, the incoherent domain is attracted to the higher-frequency region while the coherent domain is attracted to the lower-frequency region. Figure~\ref{fig:prediction-dots} clearly confirms this general result.

In summary, we have analyzed the effect of frequency heterogeneity on nonlocally coupled phase oscillators exhibiting chimera states and revealed that spatial phase locking to frequency heterogeneity occurs, i.e., the coherent domain is attracted to the low-frequency region while the incoherent domain is attracted to the high-frequency region.
We extended the self-consistency theory to the case with frequency heterogeneity and showed that it reproduces the results of numerical simulations well.
Furthermore, by using a variational argument, we explained the general mechanism leading to this behavior from the viewpoint of self-consistency.
Our results show that, by introducing weak frequency heterogeneity to chimera states of nonlocally coupled oscillators, we can softly control the position of the chimera's domains without causing large perturbations.
We expect that such dynamics can be observed in a wide range of systems with continuous translational symmetry,  including different types of oscillators and coupling schemes.
This approach could enable precise control or spatial locking of chimera states and encourage self-adaptive behaviors in various types of oscillatory networks.

	{\it Acknowledgments—  We acknowledge JSPS KAKENHI JP22K11919, JP22H00516 and JST CREST
		JP-MJCR1913 for financial support. }


\begin{thebibliography}{24}%
	\makeatletter
	\providecommand \@ifxundefined [1]{%
		\@ifx{#1\undefined}
	}%
	\providecommand \@ifnum [1]{%
		\ifnum #1\expandafter \@firstoftwo
		\else \expandafter \@secondoftwo
		\fi
	}%
	\providecommand \@ifx [1]{%
		\ifx #1\expandafter \@firstoftwo
		\else \expandafter \@secondoftwo
		\fi
	}%
	\providecommand \natexlab [1]{#1}%
	\providecommand \enquote  [1]{``#1''}%
	\providecommand \bibnamefont  [1]{#1}%
	\providecommand \bibfnamefont [1]{#1}%
	\providecommand \citenamefont [1]{#1}%
	\providecommand \href@noop [0]{\@secondoftwo}%
	\providecommand \href [0]{\begingroup \@sanitize@url \@href}%
	\providecommand \@href[1]{\@@startlink{#1}\@@href}%
	\providecommand \@@href[1]{\endgroup#1\@@endlink}%
	\providecommand \@sanitize@url [0]{\catcode `\\12\catcode `\$12\catcode
		`\&12\catcode `\#12\catcode `\^12\catcode `\_12\catcode `\%12\relax}%
	\providecommand \@@startlink[1]{}%
	\providecommand \@@endlink[0]{}%
	\providecommand \url  [0]{\begingroup\@sanitize@url \@url }%
	\providecommand \@url [1]{\endgroup\@href {#1}{\urlprefix }}%
	\providecommand \urlprefix  [0]{URL }%
	\providecommand \Eprint [0]{\href }%
	\providecommand \doibase [0]{https://doi.org/}%
	\providecommand \selectlanguage [0]{\@gobble}%
	\providecommand \bibinfo  [0]{\@secondoftwo}%
	\providecommand \bibfield  [0]{\@secondoftwo}%
	\providecommand \translation [1]{[#1]}%
	\providecommand \BibitemOpen [0]{}%
	\providecommand \bibitemStop [0]{}%
	\providecommand \bibitemNoStop [0]{.\EOS\space}%
	\providecommand \EOS [0]{\spacefactor3000\relax}%
	\providecommand \BibitemShut  [1]{\csname bibitem#1\endcsname}%
	\let\auto@bib@innerbib\@empty
	\bibitem [{\citenamefont {Kuramoto}\ and\ \citenamefont
				{Battogtokh}(2002)}]{kuramoto2002coexistence}%
	\BibitemOpen
	\bibfield  {author} {\bibinfo {author} {\bibfnamefont {Y.}~\bibnamefont
			{Kuramoto}}\ and\ \bibinfo {author} {\bibfnamefont {D.}~\bibnamefont
			{Battogtokh}},\ }\bibfield  {title} {\bibinfo {title} {Coexistence of
			coherence and incoherence in nonlocally coupled phase oscillators},\
	}\href@noop {} {\bibfield  {journal} {\bibinfo  {journal} {Nonlin. Phenom.
				Complex Syst.}\ }\textbf {\bibinfo {volume} {5}},\ \bibinfo {pages} {380}
		(\bibinfo {year} {2002})}\BibitemShut {NoStop}%
	\bibitem [{\citenamefont {Abrams}\ and\ \citenamefont
				{Strogatz}(2004)}]{abrams2004chimera}%
	\BibitemOpen
	\bibfield  {author} {\bibinfo {author} {\bibfnamefont {D.~M.}\ \bibnamefont
			{Abrams}}\ and\ \bibinfo {author} {\bibfnamefont {S.~H.}\ \bibnamefont
			{Strogatz}},\ }\bibfield  {title} {\bibinfo {title} {Chimera states for
			coupled oscillators},\ }\href@noop {} {\bibfield  {journal} {\bibinfo
			{journal} {Physical review letters}\ }\textbf {\bibinfo {volume} {93}},\
		\bibinfo {pages} {174102} (\bibinfo {year} {2004})}\BibitemShut {NoStop}%
	\bibitem [{\citenamefont {Kuramoto}(1995)}]{kuramoto1995scaling}%
	\BibitemOpen
	\bibfield  {author} {\bibinfo {author} {\bibfnamefont {Y.}~\bibnamefont
			{Kuramoto}},\ }\bibfield  {title} {\bibinfo {title} {Scaling behavior of
			turbulent oscillators with non-local interaction},\ }\href@noop {} {\bibfield
		{journal} {\bibinfo  {journal} {Progress of Theoretical Physics}\ }\textbf
		{\bibinfo {volume} {94}},\ \bibinfo {pages} {321} (\bibinfo {year}
		{1995})}\BibitemShut {NoStop}%
	\bibitem [{\citenamefont {Kuramoto}\ and\ \citenamefont
				{Nakao}(1997)}]{kuramoto1997power}%
	\BibitemOpen
	\bibfield  {author} {\bibinfo {author} {\bibfnamefont {Y.}~\bibnamefont
			{Kuramoto}}\ and\ \bibinfo {author} {\bibfnamefont {H.}~\bibnamefont
			{Nakao}},\ }\bibfield  {title} {\bibinfo {title} {Power-law spatial
			correlations and the onset of individual motions in self-oscillatory media
			with non-local coupling},\ }\href@noop {} {\bibfield  {journal} {\bibinfo
			{journal} {Physica D}\ }\textbf {\bibinfo {volume} {103}},\ \bibinfo {pages}
		{294} (\bibinfo {year} {1997})}\BibitemShut {NoStop}%
	\bibitem [{\citenamefont {Nakao}(1999)}]{nakao1999anomalous}%
	\BibitemOpen
	\bibfield  {author} {\bibinfo {author} {\bibfnamefont {H.}~\bibnamefont
			{Nakao}},\ }\bibfield  {title} {\bibinfo {title} {Anomalous spatio-temporal
			chaos in a two-dimensional system of nonlocally coupled oscillators},\
	}\href@noop {} {\bibfield  {journal} {\bibinfo  {journal} {Chaos}\ }\textbf
		{\bibinfo {volume} {9}},\ \bibinfo {pages} {902} (\bibinfo {year}
		{1999})}\BibitemShut {NoStop}%
	\bibitem [{\citenamefont {Nakagawa}\ and\ \citenamefont
				{Kuramoto}(1993)}]{nakagawa1993collective}%
	\BibitemOpen
	\bibfield  {author} {\bibinfo {author} {\bibfnamefont {N.}~\bibnamefont
			{Nakagawa}}\ and\ \bibinfo {author} {\bibfnamefont {Y.}~\bibnamefont
			{Kuramoto}},\ }\bibfield  {title} {\bibinfo {title} {Collective chaos in a
			population of globally coupled oscillators},\ }\href@noop {} {\bibfield
		{journal} {\bibinfo  {journal} {Progress of Theoretical Physics}\ }\textbf
		{\bibinfo {volume} {89}},\ \bibinfo {pages} {313} (\bibinfo {year}
		{1993})}\BibitemShut {NoStop}%
	\bibitem [{\citenamefont {Chabanol}\ \emph {et~al.}(1997)\citenamefont
				{Chabanol}, \citenamefont {Hakim},\ and\ \citenamefont
				{Rappel}}]{chabanol1997collective}%
	\BibitemOpen
	\bibfield  {author} {\bibinfo {author} {\bibfnamefont {M.-L.}\ \bibnamefont
			{Chabanol}}, \bibinfo {author} {\bibfnamefont {V.}~\bibnamefont {Hakim}},\
		and\ \bibinfo {author} {\bibfnamefont {W.-J.}\ \bibnamefont {Rappel}},\
	}\bibfield  {title} {\bibinfo {title} {Collective chaos and noise in the
			globally coupled complex ginzburg-landau equation},\ }\href@noop {}
	{\bibfield  {journal} {\bibinfo  {journal} {Physica D}\ }\textbf {\bibinfo
			{volume} {103}},\ \bibinfo {pages} {273} (\bibinfo {year}
		{1997})}\BibitemShut {NoStop}%
	\bibitem [{\citenamefont {Omel’chenko}\ \emph {et~al.}(2010)\citenamefont
				{Omel’chenko}, \citenamefont {Wolfrum},\ and\ \citenamefont
				{Maistrenko}}]{omel2010chimera}%
	\BibitemOpen
	\bibfield  {author} {\bibinfo {author} {\bibfnamefont {O.~E.}\ \bibnamefont
			{Omel’chenko}}, \bibinfo {author} {\bibfnamefont {M.}~\bibnamefont
			{Wolfrum}},\ and\ \bibinfo {author} {\bibfnamefont {Y.~L.}\ \bibnamefont
			{Maistrenko}},\ }\bibfield  {title} {\bibinfo {title} {Chimera states as
			chaotic spatiotemporal patterns},\ }\href@noop {} {\bibfield  {journal}
		{\bibinfo  {journal} {Physical Review E}\ }\textbf {\bibinfo {volume} {81}},\
		\bibinfo {pages} {065201} (\bibinfo {year} {2010})}\BibitemShut {NoStop}%
	\bibitem [{\citenamefont {Hizanidis}\ \emph {et~al.}(2014)\citenamefont
				{Hizanidis}, \citenamefont {Kanas}, \citenamefont {Bezerianos},\ and\
				\citenamefont {Bountis}}]{hizanidis2014chimera}%
	\BibitemOpen
	\bibfield  {author} {\bibinfo {author} {\bibfnamefont {J.}~\bibnamefont
			{Hizanidis}}, \bibinfo {author} {\bibfnamefont {V.~G.}\ \bibnamefont
			{Kanas}}, \bibinfo {author} {\bibfnamefont {A.}~\bibnamefont {Bezerianos}},\
		and\ \bibinfo {author} {\bibfnamefont {T.}~\bibnamefont {Bountis}},\
	}\bibfield  {title} {\bibinfo {title} {Chimera states in networks of
			nonlocally coupled hindmarsh--rose neuron models},\ }\href@noop {} {\bibfield
		{journal} {\bibinfo  {journal} {International Journal of Bifurcation and
				Chaos}\ }\textbf {\bibinfo {volume} {24}},\ \bibinfo {pages} {1450030}
		(\bibinfo {year} {2014})}\BibitemShut {NoStop}%
	\bibitem [{\citenamefont {zur Bonsen}\ \emph {et~al.}(2018)\citenamefont {zur
					Bonsen}, \citenamefont {Omelchenko}, \citenamefont {Zakharova},\ and\
				\citenamefont {Sch{\"o}ll}}]{zur2018chimera}%
	\BibitemOpen
	\bibfield  {author} {\bibinfo {author} {\bibfnamefont {A.}~\bibnamefont {zur
				Bonsen}}, \bibinfo {author} {\bibfnamefont {I.}~\bibnamefont {Omelchenko}},
		\bibinfo {author} {\bibfnamefont {A.}~\bibnamefont {Zakharova}},\ and\
		\bibinfo {author} {\bibfnamefont {E.}~\bibnamefont {Sch{\"o}ll}},\ }\bibfield
	{title} {\bibinfo {title} {Chimera states in networks of logistic maps with
			hierarchical connectivities},\ }\href@noop {} {\bibfield  {journal} {\bibinfo
			{journal} {The European Physical Journal B}\ }\textbf {\bibinfo {volume}
			{91}},\ \bibinfo {pages} {1} (\bibinfo {year} {2018})}\BibitemShut {NoStop}%
	\bibitem [{\citenamefont {Maistrenko}\ \emph {et~al.}(2015)\citenamefont
				{Maistrenko}, \citenamefont {Sudakov}, \citenamefont {Osiv},\ and\
				\citenamefont {Maistrenko}}]{maistrenko2015chimera}%
	\BibitemOpen
	\bibfield  {author} {\bibinfo {author} {\bibfnamefont {Y.}~\bibnamefont
			{Maistrenko}}, \bibinfo {author} {\bibfnamefont {O.}~\bibnamefont {Sudakov}},
		\bibinfo {author} {\bibfnamefont {O.}~\bibnamefont {Osiv}},\ and\ \bibinfo
		{author} {\bibfnamefont {V.}~\bibnamefont {Maistrenko}},\ }\bibfield  {title}
	{\bibinfo {title} {Chimera states in three dimensions},\ }\href@noop {}
	{\bibfield  {journal} {\bibinfo  {journal} {New Journal of Physics}\ }\textbf
		{\bibinfo {volume} {17}},\ \bibinfo {pages} {073037} (\bibinfo {year}
		{2015})}\BibitemShut {NoStop}%
	\bibitem [{\citenamefont {Omel'chenko}\ \emph {et~al.}(2012)\citenamefont
				{Omel'chenko}, \citenamefont {Wolfrum}, \citenamefont {Yanchuk},
				\citenamefont {Maistrenko},\ and\ \citenamefont
				{Sudakov}}]{omel2012stationary}%
	\BibitemOpen
	\bibfield  {author} {\bibinfo {author} {\bibfnamefont {O.~E.}\ \bibnamefont
			{Omel'chenko}}, \bibinfo {author} {\bibfnamefont {M.}~\bibnamefont
			{Wolfrum}}, \bibinfo {author} {\bibfnamefont {S.}~\bibnamefont {Yanchuk}},
		\bibinfo {author} {\bibfnamefont {Y.~L.}\ \bibnamefont {Maistrenko}},\ and\
		\bibinfo {author} {\bibfnamefont {O.}~\bibnamefont {Sudakov}},\ }\bibfield
	{title} {\bibinfo {title} {Stationary patterns of coherence and incoherence
			in two-dimensional arrays of non-locally-coupled phase oscillators},\
	}\href@noop {} {\bibfield  {journal} {\bibinfo  {journal} {Physical Review
				E}\ }\textbf {\bibinfo {volume} {85}},\ \bibinfo {pages} {036210} (\bibinfo
		{year} {2012})}\BibitemShut {NoStop}%
	\bibitem [{\citenamefont {Tinsley}\ \emph {et~al.}(2012)\citenamefont
				{Tinsley}, \citenamefont {Nkomo},\ and\ \citenamefont
				{Showalter}}]{tinsley2012chimera}%
	\BibitemOpen
	\bibfield  {author} {\bibinfo {author} {\bibfnamefont {M.~R.}\ \bibnamefont
			{Tinsley}}, \bibinfo {author} {\bibfnamefont {S.}~\bibnamefont {Nkomo}},\
		and\ \bibinfo {author} {\bibfnamefont {K.}~\bibnamefont {Showalter}},\
	}\bibfield  {title} {\bibinfo {title} {Chimera and phase-cluster states in
			populations of coupled chemical oscillators},\ }\href@noop {} {\bibfield
		{journal} {\bibinfo  {journal} {Nature Physics}\ }\textbf {\bibinfo {volume}
			{8}},\ \bibinfo {pages} {662} (\bibinfo {year} {2012})}\BibitemShut {NoStop}%
	\bibitem [{\citenamefont {Wickramasinghe}\ and\ \citenamefont
				{Kiss}(2013)}]{wickramasinghe2013spatially}%
	\BibitemOpen
	\bibfield  {author} {\bibinfo {author} {\bibfnamefont {M.}~\bibnamefont
			{Wickramasinghe}}\ and\ \bibinfo {author} {\bibfnamefont {I.~Z.}\
			\bibnamefont {Kiss}},\ }\bibfield  {title} {\bibinfo {title} {Spatially
			organized dynamical states in chemical oscillator networks: Synchronization,
			dynamical differentiation, and chimera patterns},\ }\href@noop {} {\bibfield
		{journal} {\bibinfo  {journal} {PloS one}\ }\textbf {\bibinfo {volume} {8}},\
		\bibinfo {pages} {e80586} (\bibinfo {year} {2013})}\BibitemShut {NoStop}%
	\bibitem [{\citenamefont {Martens}\ \emph {et~al.}(2013)\citenamefont
				{Martens}, \citenamefont {Thutupalli}, \citenamefont {Fourri{\`e}re},\ and\
				\citenamefont {Hallatschek}}]{martens2013chimera}%
	\BibitemOpen
	\bibfield  {author} {\bibinfo {author} {\bibfnamefont {E.~A.}\ \bibnamefont
			{Martens}}, \bibinfo {author} {\bibfnamefont {S.}~\bibnamefont {Thutupalli}},
		\bibinfo {author} {\bibfnamefont {A.}~\bibnamefont {Fourri{\`e}re}},\ and\
		\bibinfo {author} {\bibfnamefont {O.}~\bibnamefont {Hallatschek}},\
	}\bibfield  {title} {\bibinfo {title} {Chimera states in mechanical
			oscillator networks},\ }\href@noop {} {\bibfield  {journal} {\bibinfo
			{journal} {Proceedings of the National Academy of Sciences}\ }\textbf
		{\bibinfo {volume} {110}},\ \bibinfo {pages} {10563} (\bibinfo {year}
		{2013})}\BibitemShut {NoStop}%
	\bibitem [{\citenamefont {Kapitaniak}\ \emph {et~al.}(2014)\citenamefont
				{Kapitaniak}, \citenamefont {Kuzma}, \citenamefont {Wojewoda}, \citenamefont
				{Czolczynski},\ and\ \citenamefont {Maistrenko}}]{kapitaniak2014imperfect}%
	\BibitemOpen
	\bibfield  {author} {\bibinfo {author} {\bibfnamefont {T.}~\bibnamefont
			{Kapitaniak}}, \bibinfo {author} {\bibfnamefont {P.}~\bibnamefont {Kuzma}},
		\bibinfo {author} {\bibfnamefont {J.}~\bibnamefont {Wojewoda}}, \bibinfo
		{author} {\bibfnamefont {K.}~\bibnamefont {Czolczynski}},\ and\ \bibinfo
		{author} {\bibfnamefont {Y.}~\bibnamefont {Maistrenko}},\ }\bibfield  {title}
	{\bibinfo {title} {Imperfect chimera states for coupled pendula},\
	}\href@noop {} {\bibfield  {journal} {\bibinfo  {journal} {Scientific
				reports}\ }\textbf {\bibinfo {volume} {4}},\ \bibinfo {pages} {6379}
		(\bibinfo {year} {2014})}\BibitemShut {NoStop}%
	\bibitem [{\citenamefont {Hagerstrom}\ \emph {et~al.}(2012)\citenamefont
				{Hagerstrom}, \citenamefont {Murphy}, \citenamefont {Roy}, \citenamefont
				{H{\"o}vel}, \citenamefont {Omelchenko},\ and\ \citenamefont
				{Sch{\"o}ll}}]{hagerstrom2012experimental}%
	\BibitemOpen
	\bibfield  {author} {\bibinfo {author} {\bibfnamefont {A.~M.}\ \bibnamefont
			{Hagerstrom}}, \bibinfo {author} {\bibfnamefont {T.~E.}\ \bibnamefont
			{Murphy}}, \bibinfo {author} {\bibfnamefont {R.}~\bibnamefont {Roy}},
		\bibinfo {author} {\bibfnamefont {P.}~\bibnamefont {H{\"o}vel}}, \bibinfo
		{author} {\bibfnamefont {I.}~\bibnamefont {Omelchenko}},\ and\ \bibinfo
		{author} {\bibfnamefont {E.}~\bibnamefont {Sch{\"o}ll}},\ }\bibfield  {title}
	{\bibinfo {title} {Experimental observation of chimeras in coupled-map
			lattices},\ }\href@noop {} {\bibfield  {journal} {\bibinfo  {journal} {Nature
				Physics}\ }\textbf {\bibinfo {volume} {8}},\ \bibinfo {pages} {658} (\bibinfo
		{year} {2012})}\BibitemShut {NoStop}%
	\bibitem [{\citenamefont {Bick}\ and\ \citenamefont
				{Martens}(2015)}]{bick2015controlling}%
	\BibitemOpen
	\bibfield  {author} {\bibinfo {author} {\bibfnamefont {C.}~\bibnamefont
			{Bick}}\ and\ \bibinfo {author} {\bibfnamefont {E.~A.}\ \bibnamefont
			{Martens}},\ }\bibfield  {title} {\bibinfo {title} {Controlling chimeras},\
	}\href@noop {} {\bibfield  {journal} {\bibinfo  {journal} {New Journal of
				Physics}\ }\textbf {\bibinfo {volume} {17}},\ \bibinfo {pages} {033030}
		(\bibinfo {year} {2015})}\BibitemShut {NoStop}%
	\bibitem [{\citenamefont {Isele}\ \emph {et~al.}(2016)\citenamefont {Isele},
				\citenamefont {Hizanidis}, \citenamefont {Provata},\ and\ \citenamefont
				{H{\"o}vel}}]{isele2016controlling}%
	\BibitemOpen
	\bibfield  {author} {\bibinfo {author} {\bibfnamefont {T.}~\bibnamefont
			{Isele}}, \bibinfo {author} {\bibfnamefont {J.}~\bibnamefont {Hizanidis}},
		\bibinfo {author} {\bibfnamefont {A.}~\bibnamefont {Provata}},\ and\ \bibinfo
		{author} {\bibfnamefont {P.}~\bibnamefont {H{\"o}vel}},\ }\bibfield  {title}
	{\bibinfo {title} {Controlling chimera states: The influence of excitable
			units},\ }\href@noop {} {\bibfield  {journal} {\bibinfo  {journal} {Physical
				Review E}\ }\textbf {\bibinfo {volume} {93}},\ \bibinfo {pages} {022217}
		(\bibinfo {year} {2016})}\BibitemShut {NoStop}%
	\bibitem [{\citenamefont {Gambuzza}\ and\ \citenamefont
				{Frasca}(2016)}]{gambuzza2016pinning}%
	\BibitemOpen
	\bibfield  {author} {\bibinfo {author} {\bibfnamefont {L.~V.}\ \bibnamefont
			{Gambuzza}}\ and\ \bibinfo {author} {\bibfnamefont {M.}~\bibnamefont
			{Frasca}},\ }\bibfield  {title} {\bibinfo {title} {Pinning control of chimera
			states},\ }\href@noop {} {\bibfield  {journal} {\bibinfo  {journal} {Physical
				Review E}\ }\textbf {\bibinfo {volume} {94}},\ \bibinfo {pages} {022306}
		(\bibinfo {year} {2016})}\BibitemShut {NoStop}%
	\bibitem [{\citenamefont {Omelchenko}\ \emph {et~al.}(2016)\citenamefont
				{Omelchenko}, \citenamefont {Omel’chenko}, \citenamefont {Zakharova},
				\citenamefont {Wolfrum},\ and\ \citenamefont
				{Sch{\"o}ll}}]{omelchenko2016tweezers}%
	\BibitemOpen
	\bibfield  {author} {\bibinfo {author} {\bibfnamefont {I.}~\bibnamefont
			{Omelchenko}}, \bibinfo {author} {\bibfnamefont {O.~E.}\ \bibnamefont
			{Omel’chenko}}, \bibinfo {author} {\bibfnamefont {A.}~\bibnamefont
			{Zakharova}}, \bibinfo {author} {\bibfnamefont {M.}~\bibnamefont {Wolfrum}},\
		and\ \bibinfo {author} {\bibfnamefont {E.}~\bibnamefont {Sch{\"o}ll}},\
	}\bibfield  {title} {\bibinfo {title} {Tweezers for chimeras in small
			networks},\ }\href@noop {} {\bibfield  {journal} {\bibinfo  {journal}
			{Physical Review Letters}\ }\textbf {\bibinfo {volume} {116}},\ \bibinfo
		{pages} {114101} (\bibinfo {year} {2016})}\BibitemShut {NoStop}%
	\bibitem [{\citenamefont {Ruzzene}\ \emph {et~al.}(2019)\citenamefont
				{Ruzzene}, \citenamefont {Omelchenko}, \citenamefont {Sch{\"o}ll},
				\citenamefont {Zakharova},\ and\ \citenamefont
				{Andrzejak}}]{ruzzene2019controlling}%
	\BibitemOpen
	\bibfield  {author} {\bibinfo {author} {\bibfnamefont {G.}~\bibnamefont
			{Ruzzene}}, \bibinfo {author} {\bibfnamefont {I.}~\bibnamefont {Omelchenko}},
		\bibinfo {author} {\bibfnamefont {E.}~\bibnamefont {Sch{\"o}ll}}, \bibinfo
		{author} {\bibfnamefont {A.}~\bibnamefont {Zakharova}},\ and\ \bibinfo
		{author} {\bibfnamefont {R.~G.}\ \bibnamefont {Andrzejak}},\ }\bibfield
	{title} {\bibinfo {title} {Controlling chimera states via minimal coupling
			modification},\ }\href@noop {} {\bibfield  {journal} {\bibinfo  {journal}
			{Chaos: an interdisciplinary journal of nonlinear science}\ }\textbf
		{\bibinfo {volume} {29}} (\bibinfo {year} {2019})}\BibitemShut {NoStop}%
	\bibitem [{\citenamefont {Yao}\ \emph {et~al.}(2019)\citenamefont {Yao},
				\citenamefont {Huang}, \citenamefont {Ren}, \citenamefont {Grebogi},\ and\
				\citenamefont {Lai}}]{yao2019self}%
	\BibitemOpen
	\bibfield  {author} {\bibinfo {author} {\bibfnamefont {N.}~\bibnamefont
			{Yao}}, \bibinfo {author} {\bibfnamefont {Z.-G.}\ \bibnamefont {Huang}},
		\bibinfo {author} {\bibfnamefont {H.-P.}\ \bibnamefont {Ren}}, \bibinfo
		{author} {\bibfnamefont {C.}~\bibnamefont {Grebogi}},\ and\ \bibinfo {author}
		{\bibfnamefont {Y.-C.}\ \bibnamefont {Lai}},\ }\bibfield  {title} {\bibinfo
		{title} {Self-adaptation of chimera states},\ }\href@noop {} {\bibfield
		{journal} {\bibinfo  {journal} {Physical Review E}\ }\textbf {\bibinfo
			{volume} {99}},\ \bibinfo {pages} {010201} (\bibinfo {year}
		{2019})}\BibitemShut {NoStop}%
	\bibitem [{Note1()}]{Note1}%
	\BibitemOpen
	\bibinfo {note} {In the numerical simulation, this is replaced by a frequency
		increase at a single grid point in the system, i.e., $\omega _0 \to \omega _0
			+ \epsilon / \Delta x$ at the grid corresponding to $x = \gamma $, where
		$\Delta x = 2\pi / N$ is the grid width.}\BibitemShut {Stop}%
\end{thebibliography}
%

\end{document}